\documentclass[twocolumn,prd,a4paper,showpacs,amssymb,amsmath,nofootinbib,preprintnumbers]{revtex4}

\usepackage{graphicx} 
\usepackage{dcolumn}
\usepackage{bm}

\begin{document}

\def \d {{\rm d}}
\def \t {{\Theta}}
\def \k {{\kappa}}
\def \l {{\lambda}}

\newcommand{\beqn}{\begin{eqnarray}}
\newcommand{\eeqn}{\end{eqnarray}}
\newcommand{\AdS}{anti-de~Sitter }
\newcommand{\AAdS}{(anti-)de~Sitter }
\newcommand{\pa}{\partial}
\newcommand{\pp}{{\it pp\,}-}
\newcommand{\ba}{\begin{array}}
\newcommand{\ea}{\end{array}}

\title{Impulsive waves in the Nariai universe}

\author{Marcello Ortaggio}
 \email[Email address: ]{ortaggio@science.unitn.it}
 \affiliation{Dipartimento di Fisica, Universit\`a degli Studi di Trento, and \\ INFN, Gruppo Collegato di Trento, 38050 Povo (Trento), Italy}


\date{\today}

\begin{abstract}
A new class of exact solutions is presented which describes impulsive waves propagating in the Nariai universe. It  is constructed using a six-dimensional embedding formalism adapted to the background. Due to the topology of the latter, the wave front consists of two non-expanding spheres. Special sub-classes representing pure gravitational waves (generated by null particles with an arbitrary multipole structure) or shells of null dust are analyzed in detail. Smooth isometries of the metrics are briefly discussed. Furthermore, it is shown that the considered solutions are impulsive members of a more general family of radiative Kundt spacetimes of type-$II$. A straightforward generalization to impulsive waves in the anti-Nariai and Bertotti--Robinson backgrounds is described. For a vanishing cosmological constant and electromagnetic field, results for well known impulsive \pp waves are recovered. 

\end{abstract}


\maketitle

\section{Introduction}

\label{sec_introduction}

Many exact solutions of Einstein's equations are known which
describe radiative spacetimes (a recent review and references are
provided, e.g., in \cite{Bicak00lnp}). As special cases, impulsive waves
have been widely investigated. Their geometry is characterized by
the Dirac delta contribution to the curvature tensor, supported on a null
hypersurface, which is interpreted as an impulsive field
propagating in a given background.\footnote{As known, dealing with distributions in General 
	Relativity may lead to quantities which can not be defined within the linear Schwartz's theory. When 
	this occurs, the advanced framework of Colombeau's algebras of generalized functions is required. 		See \cite{Grosserbook}  for recent developments and applications to Einstein's theory.} 
In the simplest situation this is a constant-curvature space (Minkowski, de~Sitter or
anti-de~Sitter) and all main features of such metrics are well
known (see \cite{Podolsky02} and references therein). In particular,
these belong to two distinct families in which the waves are
either expanding or non-expanding, thus being understood as
limiting cases of sandwich waves of the Robinson--Trautman or the Kundt classes,
respectively. Specific non-expanding solutions were originally obtained by applying the Aichelburg--Sexl ultra-relativistic boost \cite{AicSex71} to different elements of the Kerr--Newman and Weyl families (see, e.g., \cite{FerPen90,LouSan92,HotTan93,BalNac95,PodGri98prd}). It has then been shown \cite{GriPod97, PodGri98cqg} that the whole class of non-expanding impulsive pure gravitational waves
(with the only exception of plane waves) is generated by null
particles with an arbitrary multipole structure, corresponding to singularities of the metric tensor. 
Moreover, as an extension of the Penrose's geometrical method \cite{Penrose72}, Dray and 't~Hooft have introduced a ``shift-function'' technique, which enables to construct non-expanding impulsive waves also in non-constant-curvature backgrounds. They used it to derive the field produced by a massless particle \cite{DratHo85npb} or by a spherical shell of
null matter \cite{DratHo85cmp} located at the horizon of a Schwarzschild black hole.
Their approach has been later generalized
\cite{Sfetsos95} to include a cosmological constant $\Lambda$ and
matter fields.

Now, a simple observation is that non-expanding impulsive waves propagating in {\em all}
possible spherically symmetric vacuum backgrounds with $\Lambda=0$ have thus been explicitly described. This is guaranteed by the celebrated Birkhoff theorem (see, e.g., \cite{kramerbook}), which leaves the Schwarzschild metric (with the Minkowski universe as a trivial sub-case) as the only possibility. However, the generalized version of this theorem admitting a non-vanishing $\Lambda$ \cite{FoyMcI72,Barnes73,kramerbook} provides a richer class of non-equivalent metrics. Namely, one has not only the Schwarzschild--\AAdS solutions, but also the Nariai metric \cite{Nariai51}, when $\Lambda>0$ (its $\Lambda<0$ counterpart, to which we shall refer as ``anti-Nariai'', admits different symmetries). 

The Nariai line-element, which actually (in a Euclidean notation) dates
back to Kasner \cite{Kasner25}, is indeed a non-singular solution
of the vacuum Einstein's equations with a positive cosmological
constant, $R_{\mu\nu}=\Lambda g_{\mu\nu}$. It is the direct
product of a two-dimensional de Sitter space with a 2-sphere. It
admits a 6-parameters group of motions and is {\em not}
conformally flat. Therefore, it is both locally and globally distinguished from the de Sitter space. Besides the
historical attention it deserved thanks to its
geometrical properties \cite{Kasner25,Nariai51,Bertotti59,kramerbook}, more recently it
has been the object of a renewed interest, since it emerges as
the extremal limit of Schwarzschild--de Sitter black holes
\cite{GinPer83,BouHaw96,CalVanZer00} (which is not equivalent to consider ``extreme'' black holes, studied, e.g., in \cite{Podolsky99grg}). Thus, it can be viewed as a
``degenerate'' black hole, in which the two
horizons have the same (maximum) size and are in thermal equilibrium at the temperature
$T=\sqrt{\Lambda}/2\pi$. Admitting a regular Euclidean section, it
turns out to be a good ``instanton'' for the study of quantum pair creation
of black holes during inflation. In more general terms, a
``charged'' version of the Nariai metric \cite{Bertotti59} (see also \cite{EliHil01}) is interpreted as a degenerate Reissner--Nordstr\"om--de Sitter black hole \cite{HawRos95}. This has been considered also in dilatonic theories, e.g. in \cite{Bousso97}.

To our knowledge, impulsive waves in the Nariai universe have not yet explicitly been studied.
It is the purpose of the present paper to construct and analyze {\em non-expanding} impulsive waves propagating in this background, thus filling a ``gap'' in the classification of impulsive waves in spherically symmetric spaces.\footnote{In Sec.~\ref{sec_curvature} it is shown how, in fact, impulsive waves in the Nariai spacetime can also be recovered as specific sub-cases of previously introduced general classes of metrics \cite{Sfetsos95,Balasin00}. The author is grateful to the referee for bringing \cite{Balasin00} to his attention.}

Sec.~\ref{sec_nariai} reviews the Nariai geometry, useful in the sequel. The complete
metric representing impulsive waves is presented in Sec.~\ref{sec_geometry}. This is done by means of a global
six-dimensional formalism adapted to the Nariai spacetime. The unusual geometry of the wave
front is described by means of six- and global natural four-coordinates. In Sec.~\ref{sec_curvature} contributions to the curvature due to the waves are calculated, and a general exact solution to the vacuum field
equations is provided. It is shown that the only possible regular solution is given by a trivial ``gauge'' term, which can be removed by an appropriate coordinate transformation. Then, the unavoidable singularities are interpreted as point sources of pure gravitational waves. 
Also, a complementary situation is discussed in
which there is no impulse in the Weyl scalars and the
gravitational field is entirely due to shells of null matter.
Sec.~\ref{sec_symmetries} deals with smooth symmetries of the
 impulsive metrics. In Sec.~\ref{sec_radiative} it is shown that these non-expanding impulsive waves are in fact a limiting case of a more general type-$II$ spacetime of the Kundt class, as a ``profile'' function approaches the Dirac delta. It is thus suggested to interpret this general spacetime as the Nariai universe with gravitational radiation. A straightforward generalization to impulsive waves in other direct product
backgrounds, in particular the Bertotti--Robinson universe, is sketched in Sec.~\ref{sec_directproducts}.

\section{The Nariai spacetime}

\label{sec_nariai}

The Nariai universe \cite{Nariai51,Kasner25} can be conveniently visualized as a 4-submanifold of a six-dimensional Lorentzian flat manifold 
\begin{equation}
\d s^2= -\d {Z_0}^2+\d {Z_1}^2+\d{Z_2}^2+\d{Z_3}^2+\d{Z_4}^2+\d{Z_5}^2 ,
\label{general}
\end{equation}
determined by two constraints
\begin{equation}
 -{Z_0}^2+{Z_1}^2+{Z_2}^2=a^2 , \qquad   {Z_3}^2+{Z_4}^2+{Z_5}^2=a^2 ,
 \label{constraints}
\end{equation}
where $a>0$ is related to the cosmological constant by 
\begin{equation}
\Lambda=\frac{1}{a^2} .
\end{equation} 
It is then obvious that the spacetime is the direct product dS$_2\times {\mathbb S}^2$ of two constant curvature 2-spaces, thus being ``symmetric'' (i.e., $R_{\mu\nu\rho\sigma;\tau}=0$) \cite{kramerbook}, and admits a {\em six-dimensional group of isometries} $SO(2,1)\times SO(3)$. In particular, it is spherically symmetric (but not isotropic) and (locally) static.
Furthermore, the group acts {\em transitively} (i.e., the spacetime is homogeneous) and has a two-dimensional isotropy subgroup at each point, composed of one boost and one spatial rotation. Note that the charged Nariai solution \cite{Bertotti59} is obtained by replacing the second constraint in (\ref{constraints}) with ${Z_3}^2+{Z_4}^2+{Z_5}^2=b^2$  (where $b=\mbox{constant}\neq a$), provided $a^{-2}+b^{-2}=2\Lambda$.

Various four-dimensional parametrizations of (\ref{general}) with (\ref{constraints}) are known. The static
Schwarzschild-like one (covering only a part of the whole manifold)
\beqn
 \d s^2= & - & \left(1-\frac{r^2}{a^2}\right)\d t^2+\left(1-\frac{r^2}{a^2}\right)^{-1}\d r^2 \nonumber \\
	 & + & a^2(\d\theta^2+\sin^2\theta\d\phi^2) ,
 \label{dSform}
\eeqn
is given by (for $0<r<a$)
\beqn
  Z_0 & = & \sqrt{a^2-r^2}\sinh(t/a) , \quad Z_1=\sqrt{a^2-r^2}\cosh(t/a) , \nonumber \\
  Z_2 & = & r , \qquad Z_3=a\sin\theta\cos\phi , \\
  Z_4 & = & a\sin\theta\sin\phi, \qquad Z_5=a\cos\theta \nonumber .
 \label{dScoords}
\eeqn

\begin{figure}
 \includegraphics[width=7.8cm]{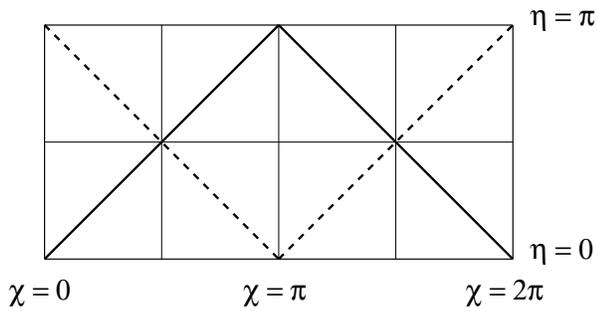}
 \caption{The conformal diagram of the non-singular Nariai universe in coordinates $(\eta,\chi)$ of 
	(\ref{global}) and (\ref{conftime}). The spacelike boundaries $\eta=0$ and $\eta=\pi$ correspond to
	$\tau=-\infty$ and $\tau=+\infty$, respectively. The angular coordinate $\chi$ spans a circle, so that 
	$\chi=0$ and $\chi=2\pi$ are identified. Each point of this representation is a 2-sphere in the actual 
	spacetime. Solid and dashed null lines represent the (disconnected) past and future event horizon, 
	respectively, for a geodesic observer $\chi=0$.}
 \label{fig_conformal}
\end{figure}

With the natural re-definition
\beqn
 Z_0 & = & a\sinh(\tau/a) , \nonumber \\
 Z_1 & = & a\cosh(\tau/a)\cos\chi , \\
 Z_2 & = & a\cosh(\tau/a)\sin\chi \nonumber ,
\eeqn
for $\tau\in(-\infty,+\infty)$ and $\chi\in[0,2\pi]$ periodic, one gets the global Kantowski--Sachs cosmological line-element
\begin{equation}
 \d s^2=-\d\tau^2+a^2\cosh^2(\tau/a)\d \chi^2+a^2(\d\theta^2+\sin^2\theta\d\phi^2) ,
 \label{global}
\end{equation}
in which the ${\mathbb R}\times {\mathbb S}^1\times {\mathbb S}^2$ topology is manifest. Note that while the ${\mathbb S}^1$ factor describes a circle which shrinks to a minimum radius $a$ at $\tau=0$ and then re-expands, the ${\mathbb S}^2$ has a constant radius $a$ at any time (this is different from the well known behaviour of the de Sitter universe, whose ${\mathbb S}^3$ spatial section contracts and expands isotropically). 
Now, for constant $\theta$ and $\phi$, it is straightforward to visualize the conformal structure of the spacetime by defining a conformal time 
\begin{equation}
 \eta=2\arctan(e^{\tau/a})\in[0,\pi] .
 \label{conftime}
\end{equation}
The diagram (Fig.~\ref{fig_conformal}) is that of a two-dimensional de Sitter space, with a spacelike infinity for timelike and null lines.
Obviously, each point of the diagram corresponds to a two-sphere of constant radius $a$, parametrized by $(\theta,\phi)$. For the geodesic observer $\chi=0$, for instance, both the future and past event horizon consist of two connected components. These are given for the former by $\eta_+^f=\chi-\pi$ and $\eta_-^f=\pi-\chi$, and for the latter by $\eta_+^p=\chi$ and $\eta_-^p=2\pi-\chi$. 

In order to express the curvature tensor, we introduce another suitable coordinate system. First, we define six-dimensional null coordinates $U=\frac{1}{\sqrt2}(Z_0+Z_1)$ and $V=\frac{1}{\sqrt2}(Z_0-Z_1)$, and then 
\beqn
 & & U=\frac{u}{\Omega} , \qquad V=-\frac{v}{\Omega} , \qquad Z_2=\frac{1-\Lambda uv}{\sqrt{2\Lambda}\Omega} , \nonumber  
 \label{nullcoord} \\
 & & Z_3=\frac{\zeta+\bar{\zeta}}{\sqrt{2}\Sigma} , \quad Z_4=-i\frac{\zeta-\bar{\zeta}}{\sqrt{2}\Sigma} , \quad Z_5=\frac{2-\Sigma}{\sqrt{\Lambda}\Sigma} ,
\eeqn
with
\begin{equation}
 \Omega={\textstyle\frac{1}{\sqrt{2}}}(1+\Lambda uv), \qquad \Sigma=1+{\textstyle\frac{1}{2}}\Lambda\zeta\bar{\zeta} .
\end{equation}
This gives a Kruskal form
\begin{equation}
 \d s^2=\frac{4\d u\d v}{(1+\Lambda uv)^2}+\frac{2\d\zeta\d\bar{\zeta}}{(1+\frac{1}{2}\Lambda\zeta\bar{\zeta})^2} ,
 \label{kruskal}
\end{equation} 
in which the limit $\Lambda\rightarrow 0$ can be explicitly performed, leading to the Minkowski spacetime.
Using the natural null tetrad $\mbox{\boldmath$k$}=\Omega\,\pa_v$, $\mbox{\boldmath$l$}=-\Omega\,\pa_u$, $\mbox{\boldmath$m$}=\Sigma\pa_{\bar{\zeta}}$, the only non-trivial curvature components are
\begin{equation}
 \Psi_2=-\frac{\Lambda}{3} , \qquad R=4\Lambda .
 \label{curvature_nariai}
\end{equation}
This explicitly demonstrates that the Nariai spacetime is a Petrov type-$D$ solution of vacuum Einstein's equations with a positive cosmological constant. 

\section{Geometry of impulsive waves}

\label{sec_geometry}

\begin{figure}
 \includegraphics[width=7cm]{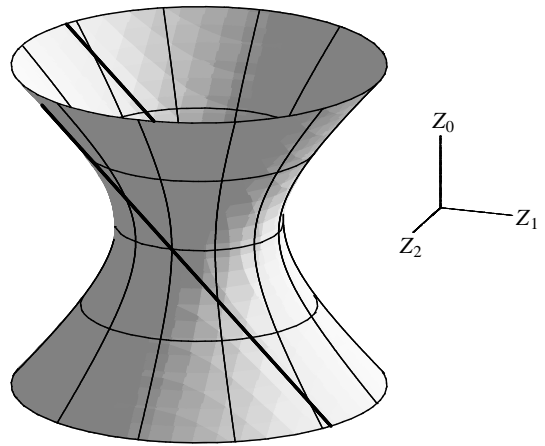}
 \caption{The 2-hyperboloid visualizes the dS$_2$ factor of the Nariai spacetime dS$_2\times {\mathbb S}^2$, once that the coordinates $Z_3$, $Z_4$ and $Z_5$ have been suppressed (see (\ref{general}) and (\ref{constraints})). Then, each point corresponds to a 2-sphere of a constant area $4\pi a^2$ in the four-dimensional spacetime. The parallel straight lines $Z_0+Z_1=0$ are the histories of two of these spheres, which propagate at the speed of light and represent the impulse (\ref{impsurface}).}
 \label{fig_hyperboloid}
\end{figure}

\begin{figure}
 \includegraphics[width=8.4cm]{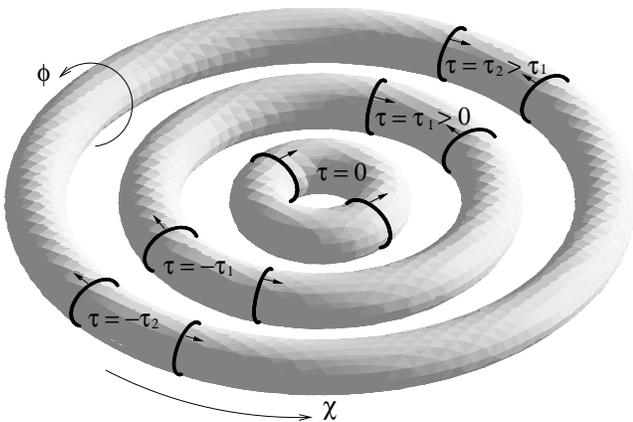}
 \caption{Suppressing the $\theta$ in global coordinates (\ref{global}), each spatial section of the Nariai universe reduces to a 2-torus parametrized by $(\chi,\phi)$, as drawn for different values of the proper time $\tau$.  The torus is contracting for $\tau<0$, reaches its minimum size at $\tau=0$, and then re-expands indefinitely as $\tau\rightarrow+\infty$. However, such an expansion is anisotropic, only involving the angular coordinate $\chi$. At any time, the impulsive wave front consists of two non-expanding 2-spheres, represented here by circles of constant radius. These propagate in opposite directions from one side of the universe ($\chi=0$) to the other ($\chi=\pi$), as $\tau$ grows from $-\infty$ to $+\infty$.}
 \label{fig_tori}
\end{figure}

It is known that non-expanding impulsive waves in the \AAdS backgrounds can be conveniently described as five-dimensional impulsive {\it pp\,}-waves plus an appropriate constraint \cite{HotTan93,PodGri98cqg}. In close analogy, we introduce here a class of impulsive waves propagating in the Nariai universe as a six-dimensional {\it pp\,}-wave constrained by (\ref{constraints}), i.e. as a metric
\beqn
 \d s^2= & - & 2\d U\d V+\d{Z_2}^2+\d{Z_3}^2+\d{Z_4}^2+\d{Z_5}^2 \nonumber  \label{6-wave}
\\ 
 	& + & \tilde{H}(Z_2,Z_3,Z_4,Z_5)\,\delta(U)\,\d U^2 ,
\eeqn
with
\begin{equation}
 -2UV+{Z_2}^2=a^2 , \qquad {Z_3}^2+{Z_4}^2+{Z_5}^2=a^2 ,
 \label{nullconstraints}
\end{equation}
where $\delta(U)$ is the Dirac distribution. For $U\neq 0$ the spacetime (\ref{6-wave}) (with (\ref{nullconstraints})) obviously reduces to the Nariai background. The impulse is located on the null 3-manifold $U=0=Z_0+Z_1$, given by
\begin{equation}
 Z_2=\pm a , \qquad {Z_3}^2+{Z_4}^2+{Z_5}^2=a^2 .
 \label{impsurface}
\end{equation}
This is the history of two non-intersecting and {\em non-expanding} 2-spheres of constant area $4\pi a^2$, so that the spatial sections of the 3-wave front are disconnected 2-manifolds (see Fig.~\ref{fig_hyperboloid}).
In the global coordinates of (\ref{global}), which display the ${\mathbb S}^1\times{\mathbb S}^2$ spatial sections of the universe, the impulsive wave stays at $\cos\chi=-\tanh(\tau/a)$, with its connected components at 
\beqn
 \chi_+ & = & \arccos[-\tanh(\tau/a)] , \nonumber  \label{impglobal} \\
 \chi_- & = & 2\pi-\arccos[-\tanh(\tau/a)] ,
\eeqn
respectively. These are nothing but the components $\eta_{\pm}^p$ of the past event horizon of the geodesic observer $\chi=0$ in the Nariai cosmos, see Sec.~\ref{sec_nariai} and Fig.~\ref{fig_conformal}. Formula (\ref{impglobal}) demonstrates that the two ${\mathbb S}^2$-components propagate in opposite directions along the circle ${\mathbb S}^1$, from $\chi_+=0$, $\chi_-=2\pi$ as $\tau\rightarrow-\infty$, to $\chi_+=\pi/2$, $\chi_-=3\pi/2$  at $\tau=0$ and $\chi_{\pm}=\pi$ as $\tau\rightarrow+\infty$. Thus, the history of each of them spans exactly one-half of the ${\mathbb S}^1$. Note that, although their $\chi$-separation decreases (as $\tau\rightarrow\pm\infty$), they never collide. Indeed, the ${\mathbb S}^1$ contracts and then re-expands in such a way that the spatial separation between these, $\Delta l=a\cosh(\tau/a)[(\chi_{-}-\chi_+)\mod{2\pi}]$, is always finite. In particular, this reaches its maximum value $\pi a$ at $\tau=0$, when the circle contracts to a minimum radius $a$, and approaches $2a$ in the limit $\tau\rightarrow\pm\infty$, when the circle expands indefinitely. In order to visualize the propagation of the impulsive wave, let us suppress one spatial dimension by fixing  $\theta=\theta_0=\mbox{constant}\neq 0,\, \pi$ in (\ref{global}). Then, each connected component of the wave front reduces to a circle of a constant radius $a\sin\theta_0$, spanned by $\phi$, which propagates on a flat 2-torus ${\mathbb S}^1\times{\mathbb S}^1$, with coordinates $(\chi,\phi)$. This is represented in Fig.~\ref{fig_tori}  (which is not completely faithful, as a flat 2-torus can not be isometrically immersed in ${\mathbb R}^3$).
It is interesting to compare the description of the present geometry and its global structure with that of non-expanding waves in a de~Sitter spacetime, given in \cite{PodGri97}. 

\section{Curvature and Einstein's equations}

\label{sec_curvature}

In this section we evaluate the curvature tensor and solve the vacuum equations associated with the impulsive waves presented in Sec.~\ref{sec_geometry}. 
With the parametrization (\ref{nullcoord}) (and a trivial re-scaling $H\equiv\sqrt{2}\tilde{H}$) the metric (\ref{6-wave}) with (\ref{nullconstraints}) becomes  
\begin{equation}
 \d s^2=\frac{H(\zeta,\bar{\zeta})\delta(u)\d u^2+4\d u\d v}{(1+\Lambda uv)^2}+\frac{2\d\zeta\d\bar{\zeta}}{(1+\frac{1}{2}\Lambda\zeta\bar{\zeta})^2} .
 \label{4-wave}
\end{equation}
Note that (\ref{4-wave}) could equivalently be obtained by means of the Dray--'t~Hooft shift-function method \cite{DratHo85npb,Sfetsos95} applied to~(\ref{kruskal}), or by using the general construction based on the generalized Kerr--Schild class \cite{Balasin00}.\footnote{\label{note_KS}The metric (\ref{4-wave}) is indeed naturally decomposed as $g_{\mu\nu}={\textstyle\frac{1}{2}}H\delta(u)k_\mu k_\nu+\tilde{g}_{\mu\nu}$, where $\tilde{g}_{\mu\nu}$ corresponds to the Nariai background (\ref{kruskal}).}
Now, we employ the null tetrads formalism. We replace $\mbox{\boldmath$l$}$ in the null tetrad of Sec.~\ref{sec_nariai} with the more general $\mbox{\boldmath$l$}=\Omega\left(-\pa_u+\frac{1}{4}H\delta(u)\pa_v\right)$, and consider the distributional identity $u\delta(u)=0$. Expressions (\ref{curvature_nariai}) remain unchanged and the only new curvature components are
\beqn
 \Psi_4 & = & -{\textstyle\frac{1}{4}}\Sigma\left(\Lambda\bar{\zeta}H_\zeta+\Sigma H_{\zeta\zeta}\right)\delta(u)  , \nonumber \\
 \Phi_{22} & = & -{\textstyle\frac{1}{4}}\Sigma^2H_{\zeta\bar{\zeta}}\,\delta(u)  .
 \label{NPcurvature}
\eeqn
Thus, in general the metric (\ref{4-wave}) describes impulsive gravitational waves plus an impulse of null matter localized at $u=0$. Correspondingly, on the wavefront the spacetime is of Petrov type-$II$, with energy-momentum representing pure radiation ($4\pi\,T_{\mu\nu}=\Phi_{22}\,k_{\mu}k_{\nu}$).
The impulsive contribution to the Weyl tensor is of type-$N$ (as expected from the general theory of lightlike shells \cite{BarIsr91}, see also \cite{BarBreHog97,BarHog98}). The limit $\Lambda\rightarrow 0$ in (\ref{4-wave}) and (\ref{NPcurvature}) leads to results for well known impulsive \pp waves.

{\em Pure gravitational waves} occur when the simple vacuum field equation $H_{\zeta\bar{\zeta}}=0$ is satisfied, i.e. for
\begin{equation}
 H(\zeta,\bar{\zeta})=f(\zeta)+\bar{f}(\bar{\zeta}) ,
 \label{vacuumsol}
\end{equation}
in which $f(\zeta)$ is an arbitrary analytic function of $\zeta$. This is again formally analogous to well known {\em pp\,}-waves \cite{kramerbook,GriPod97}, except that the coordinates $(\zeta,\bar{\zeta})$ span 2-spheres, here, instead of 2-planes. In particular, for $H=a_0=\mbox{constant}$ the line-element (\ref{4-wave}) represents only the Nariai background in different coordinates, since the impulse is removable by the discontinuous
coordinate transformation
\begin{equation}
 u'=\frac{u}{1-{\textstyle\frac{1}{4}}a_0\Lambda u\t(u)} , \qquad v'=v+{\textstyle\frac{1}{4}}a_0\t(u) , \qquad  \zeta'=\zeta , 
\end{equation}
where $\t(u)$ is the step function. This result is in full agreement with the Birkhoff theorem, as a metric (\ref{4-wave}) with a constant $H$ clearly represents a spherically symmetric vacuum spacetime.

General non-trivial solutions (\ref{vacuumsol}) necessarily
contain singularities localized on the wavefront, which can be
considered as null point sources of impulsive gravitational waves. In order
to achieve such a physical interpretation, it is convenient to
follow \cite{PodGri98cqg} in introducing a coordinate
\begin{equation}
 z=\cos\theta ,
\end{equation} so that $Z_3=a\sqrt{1-z^2}\cos{\phi}$,
$Z_4=a\sqrt{1-z^2}\sin{\phi}$, $Z_5=az$. This parametrization
enables us to rewrite the matter content of the spacetime as
\begin{equation}
\Phi_{22}=-{\textstyle\frac{1}{8}}{\bf \Delta} H\,\delta(u) , \end{equation} where
${\bf \Delta}\equiv
\Lambda\{\pa_z[(1-z^2)\pa_z]+(1-z^2)^{-1}\pa_\phi\pa_\phi\}$ is the
Laplacian on a 2-sphere. By the standard method, one can
now solve the vacuum equation $\Phi_{22}=0$ separating
$H(z,\phi)={\cal Z}(z)\Phi(\phi)$. To each angular 
mode $\Phi_m=\cos[m(\phi-\phi_m)]$ (with $m=0,1,2,\ldots$ and
$\phi_m$ being arbitrary ``phase'' constants) corresponds an associated
Legendre equation $\{\pa_z[(1-z^2)\pa_z]-m^2/(1-z^2)\}{\cal
Z}_m=0$ (with missing $l(l+1)$-term). For each value of $m$, this
has the general solutions
\beqn
 & & {\cal Z}_0(z)=a_0+\frac{b_0}{2}\ln\frac{1+z}{1-z}
 \nonumber  , \\
 & & {\cal Z}_m(z)=b_mF_m(z)+b_{-m}F_{-m}(z) \quad (m\ge 1) ,
\eeqn where $a_0$, $b_0$ and $b_{\pm m}$ are ``amplitude'' constants
and $F_{\pm m}(z)$ are defined by the recurrence formula
\begin{equation}
 F_{\pm m}(z)\equiv (1-z^2)^{m/2}\frac{\d^m}{\d z^m}\ln(1\mp z)^{1/2} .
 \label{Fm}
\end{equation} 
These functions, which are basically positive and negative
powers of $[(1+z)/(1-z)]^{1/2}=\cot(\theta/2)$, are singular at $z=\pm 1$, respectively. The
general vacuum solution is then given by a superposition
\beqn
  & & H(z,\phi)=a_0+\frac{b_0}{2}\ln\frac{1+z}{1-z} \nonumber \\
  & & {}+\sum_{m=1}^\infty[b_mF_m(z)+ b_{-m}F_{-m}(z)]\cos[m(\phi-\phi_m)] . \qquad
 \label{purewaves}
\eeqn
Recalling that $a_0$ represents a removable term, it is now
clear that non-trivial solutions (\ref{purewaves}) contain at
least one singularity at $z=1$ or $z=-1$, i.e. at one of the poles
$\theta=0,\,\pi$ of the ``twin'' spherical wave surfaces. By defining the
source term as $J(z,\phi)\equiv -a^2{\bf \Delta} H(z,\phi)$, one finds
$J(z,\phi)=b_0J_0(z)+\sum_{m=1}^\infty[b_mJ_m(z,\phi)+b_{-m}J_{-m}(z,\phi)]$.
The $m$-components are given by
\begin{align}
 & J_0(z)=\delta(1-z)-\delta(1+z) , \nonumber \\
 & J_{\pm m}(z,\phi)=-(1-z^2)^{m/2}\delta^{(m)}(1\mp z)\cos[m(\phi-\phi_m)] ,
 \label{mpole}
\end{align}
where $\delta^{(m)}$ is the $m$-derivative of $\delta$. 
Calculations which justify (\ref{mpole}) follow the approach
of \cite{PodGri98cqg} and are summarized in Appendix~\ref{app_multipoles}.
According to (\ref{mpole}), for $m\neq 0$ each $J_{\pm m}$ term 
in the energy-momentum tensor describes {\em a single point source with an
$m$-pole structure}. For $m=0$, instead, there is {\em a pair of ``monopole'' particles} (compare with \cite{GriPod97, PodGri98cqg}). Thus, the general solution (\ref{purewaves}) contains null particles at the
poles of both twin 2-spheres which compose the wave
front. However, $\Phi_{22}$ is linear in $H$ and the background is
invariant under rotations. Hence, in general, one can superimpose
any number of such arbitrary multipole particles
arbitrarily located over the impulsive surfaces. Note that the monopole term
in $b_0$ describes an axially symmetric
spacetime. This is the counterpart of the
Aichelburg--Sexl \cite{AicSex71} and Hotta--Tanaka \cite{HotTan93}
solutions for impulsive waves in constant curvature spaces. A comment on the energy conditions satisfied by these sources is given in Appendix~\ref{app_energy}.

It is finally natural to investigate the complementary situation in which there is no gravitational impulse in the Weyl scalars. Again using the coordinates $(z,\phi)$ on the wave front, the complex equation $\Psi_4=0$ splits into its real and imaginary parts
\begin{equation} 
 (1-z^2)^2H_{zz}-H_{\phi\phi}=0 , \qquad (1-z^2)H_{z\phi}+zH_\phi=0 .
\end{equation}
After separation of variables, these have the general solution (up to a removable constant term)
\begin{equation}
  H(z,\phi)=b_0z+b_1\sqrt{1-z^2}\cos(\phi-\phi_1) .
 \label{pureshell}
\end{equation}
In this case, the gravitational field is entirely generated by two spherical {\em shells of null matter} which form the impulse. In particular, no point particles (singularities) appear. Again, $b_0$ is the coefficient of an axially symmetric term.

Except for the ``pure'' solutions (\ref{purewaves}) and (\ref{pureshell}), the spacetime (\ref{4-wave}) in general describes impulsive waves generated by an arbitrary superposition of multipole point-like sources and an impulse of null dust in the Nariai universe.

\section{Symmetries of the impulsive solutions}

\label{sec_symmetries}

In \cite{PodOrt01} (smooth) symmetries of non-expanding impulsive waves in \AAdS spacetime have been investigated using an embedding formalism similar to that of (\ref{6-wave}) and (\ref{nullconstraints}). In particular, it has been demonstrated that these are the transformations which leave both the four-background and the embedding space (a five-dimensional {\em pp\,}-wave) unchanged.

An analogous analysis can be performed in the present case. Again, {\em smooth} symmetries of the full spacetime must also be symmetries of the background (described in Sec.~\ref{sec_nariai}). Among these, it is natural to consider the isometries of the embedding space (\ref{6-wave}) (see \cite{AicBal96}). Then, it turns out that spacetimes representing non-expanding impulsive waves in the Nariai universe {\em in general} admit at least one Killing vector field. Namely, for an arbitrary $\tilde{H}$ the metric (\ref{6-wave}) and the constraints (\ref{nullconstraints}) are invariant under the null rotation generated by
\begin{equation}
 Z_2\,\pa_V+ U\,\pa_{Z_2} ,
\end{equation}
i.e. under the transformation ($\tilde{\beta}$\ being a parameter)
\begin{eqnarray}
 & U' & =U , \qquad V'= V+\tilde{\beta}\,Z_2+{\textstyle\frac{1}{2}} \tilde{\beta}^2\, U , \nonumber \\
 & Z_2' & =Z_2+\tilde{\beta}\, U , \qquad Z_i'=Z_i \quad (i=3,4,5) .
\end{eqnarray}
In terms of the four-coordinates of (\ref{4-wave}), this corresponds to the generator
\begin{equation}
 \pa_v+\Lambda u^2\pa_u ,
\end{equation}
with the finite transformation
\begin{equation}
 u'=\frac{u}{1-\beta\Lambda u} , \qquad  v'=v+\beta , \qquad \zeta'=\zeta .
\end{equation}
In the limit $\Lambda\rightarrow 0$, this simply becomes the translation of \pp waves, generated by $\pa_v$. Particular choices of $\tilde{H}$ may represent spacetimes with more isometries. For instance, in the case $\tilde{H}=\tilde{H}(Z_5)$ one has an axially symmetric metric independent of $\phi$, with the further Killing vector $Z_3\pa_{Z_4}-Z_4\pa_{Z_3}$ (see Sec.~\ref{sec_curvature} for two explicit examples). 
From the Killing equations it can be easily verified that, in the axially symmetric case, no more symmetries are permitted (except for a trivial $\tilde{H}$).
A richer structure is in principle expected if one allows for non-smooth transformations, for specific shapes of the profile function $\tilde{H}$. However, we do not deal with this issue here, as it involves mathematical subtleties which go beyond the scope of this paper (see \cite{AicBal97} for such a study in the case of impulsive {\em pp\,}-waves).

\section{The limit of exact sandwich waves}

\label{sec_radiative}

In this section we wish to demonstrate that non-expanding impulsive waves in the Nariai spacetime constructed above can be naturally understood as limiting cases of more general exact radiative spacetimes.
Within the wide Kundt class of non-diverging solutions \cite{kramerbook}, let us concentrate here on the sub-family
\begin{equation}
 \d s^2=\d u^2\left(\Lambda w^2+2H\right)-2\d u\d w+\frac{2\d\zeta\d\bar{\zeta}}{(1+\frac{1}{2}\Lambda\zeta\bar{\zeta})^2} ,
  \label{kundt}
\end{equation}
in which $H=H(u,\zeta,\bar{\zeta})$ is analytic in $(\zeta,\bar{\zeta})$ and has an arbitrary dependence on $u$.
With the standard null tetrad $\mbox{\boldmath$k$}=\pa_w$, $\mbox{\boldmath$l$}=\pa_u+\frac{1}{2}(\Lambda w^2+2H)\pa_w$, $\mbox{\boldmath$m$}=\Sigma\pa_{\bar{\zeta}}$, the curvature components read
\beqn
 \Psi_2 & = & -\frac{\Lambda}{3} , \qquad \Psi_4=-\Sigma\left(\Lambda\bar{\zeta}H_\zeta+\Sigma H_{\zeta\zeta}\right)  , \nonumber \\
 \Phi_{22} & = & -\Sigma^2H_{\zeta\bar{\zeta}} , \qquad R=4\Lambda .
\eeqn
Thus, in general the spacetime is of type-$II$. The vacuum equation $\Phi_{22}=0$ clearly has the solution
\begin{equation}
 H(u,\zeta,\bar{\zeta})=f(u,\zeta)+\bar{f}(u,\bar{\zeta}) ,
 \label{vacuumsol2}
\end{equation}
see the discussion in Sec.~\ref{sec_curvature}. In this case, the metric (\ref{kundt}) represents the Kundt vacuum spacetime with a positive cosmological constant. In particular, for $H=0$ this turns out to be the Nariai universe (\ref{kruskal}), as the substitution
\begin{equation}
  w=-\frac{\sqrt{2}}{\Omega}v 
 \label{kundtcoord}
\end{equation}
explicitly shows. 

Now, we can understand the null coordinate $u$ in (\ref{kundt}) as playing the role of a ``retarded time'' and the function $H$ that of a ``wave profile''. Then, if $H$ is taken to be non-vanishing only for a finite range of $u$, it is natural to interpret the metric (\ref{kundt})  with (\ref{vacuumsol2}) as describing a pure gravitational field of finite duration which propagate in a Nariai background (with the speed of light). Moreover, using (\ref{kundtcoord}) it is easy to verify that in the limit of ``instantaneous'' waves, $H(u,\zeta,\bar{\zeta})\rightarrow H(\zeta,\bar{\zeta})\delta(u)$, spacetimes (\ref{kundt}) are exactly the previously considered non-expanding impulsive waves (\ref{4-wave}). This is a strict analogy with non-expanding impulsive waves in constant curvature backgrounds, which are known to be impulsive members of the Kundt class of type-$N$ solutions of vacuum Einstein's equations (see \cite{Podolsky98nonexp}). In general, one can construct sandwich gravitational waves with an {\em arbitrary} profile, e.g. shock or smooth waves. In any case, these waves are spherical but non-expanding.

In view of these results, we suggest to interpret the exact solutions (\ref{kundt}) (with (\ref{vacuumsol2})) as describing a Nariai universe containing gravitational radiation. 

\section{Impulsive waves in other direct product backgrounds}

\label{sec_directproducts}

\begin{figure}
 \includegraphics[width=4.6cm]{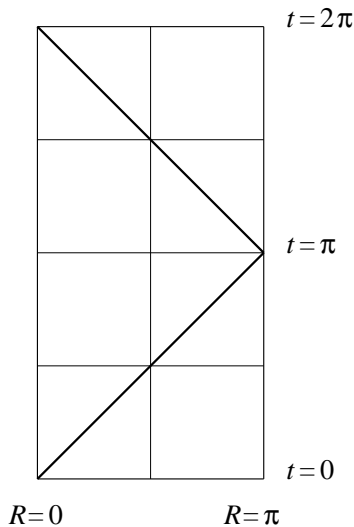}
 \caption{The conformal diagram of the Bertotti--Robinson and anti-Nariai spacetimes with non-expanding impulsive waves, given by the two null lines. Each point represents a two-dimensional (pseudo-)sphere. The timelike boundaries $R=0$ and $R=\pi$ correspond to null and spacelike infinity on opposite sides of the universe. As long as the coordinate time $t$ is assumed to be periodic, $t=0$ and $t=2\pi$ have to be identified. Otherwise, one can unwrap $t$ and build an endless tower of these conformal diagrams, from $t=-\infty$ to $t=+\infty$.}
 \label{fig_anticonformal}
\end{figure}

The construction of Sec.~\ref{sec_geometry} can be easily generalized to describe non-expanding impulsive waves propagating in other well known spacetimes which are the direct product of two 2-spaces with non-vanishing constant curvature. Namely, we can replace (\ref{6-wave}) and (\ref{nullconstraints}) with the more general metric
\beqn
  \d s^2= & - & 2\d U\d V+\epsilon_1\d{Z_2}^2+\d{Z_3}^2+\d{Z_4}^2+\epsilon_2\d{Z_5}^2 \nonumber \\
	  & + & \tilde{H}(Z_2,Z_3,Z_4,Z_5)\,\delta(U)\,\d U^2 ,
\eeqn
\begin{equation}
   -2UV+\epsilon_1 {Z_2}^2=\epsilon_1 a^2 , \quad {Z_3}^2+{Z_4}^2+\epsilon_2{Z_5}^2=\epsilon_2 a^2 ,
\end{equation}
in which $\epsilon_1,\,\epsilon_2=\pm 1$ give the sign of the curvature of each 2-space. A four-parametrization which generalizes (\ref{nullcoord}) and (\ref{4-wave}) reads
\begin{equation}
 \d s^2=\frac{H(\zeta,\bar{\zeta})\delta(u)\d u^2+4\d u\d v}{(1+\epsilon_1a^{-2}uv)^2}+\frac{2\d\zeta\d\bar{\zeta}}{(1+\frac{1}{2}\epsilon_2a^{-2}\zeta\bar{\zeta})^2} .
 \label{generalwave}
\end{equation}
We are thus left with four possible different non-trivial spacetimes, according to the signs of $\epsilon_1$ and $\epsilon_2$. Note that backgrounds with $\epsilon_1=-1$ contain closed timelike curves, unless one takes a universal covering. The case $\epsilon_1=\epsilon_2=+1$ corresponds to the impulsive solution in the Nariai background described in previous sections (with $\Lambda=a^{-2}$). When $\epsilon_1=\epsilon_2=-1$, one has waves in the anti-Nariai spacetime AdS$_2\times{\mathbb H}^{\,2}$. This is a vacuum solution with $\Lambda=-a^{-2}<0$, which appears, for example, in the extremal limit of topological black holes \cite{CalVanZer00} (in which case ${\mathbb H}^{\,2}$ is compactified to a Riemann surface of genus $g>1$). The famous Bertotti--Robinson metric AdS$_2\times{\mathbb S}^2$ \cite{Bertotti59,Robinson59} is recovered when $\epsilon_1=-\epsilon_2=-1$, and describes a conformally flat spacetime filled with a uniform electromagnetic field. The last possibility, $\epsilon_1=-\epsilon_2=1$, gives a conformally flat ``unphysical'' spacetime dS$_2\times{\mathbb H}^{\,2}$ with negative energy density. Combined situations ($\Lambda$ plus an electromagnetic field) may also occur by applying the prescription suggested in Sec.~\ref{sec_nariai} for the charged Nariai solution, see \cite{Bertotti59}. In any case, metrics (\ref{generalwave}) in general describe a superposition of impulsive gravitational waves (with point-like sources) and impulsive null dust, the geometry of the wave front being ${\mathbb S}^2$ (for $\epsilon_2=+1$) or ${\mathbb H}^{\,2}$ (for $\epsilon_2=-1$). The essential conformal structure  depends only on the first factor in the direct product metric, which contains $\epsilon_1$. For the dS$_2$ is shown in Fig.~\ref{fig_conformal}. For the AdS$_2$, one similarly obtains that of Fig.~\ref{fig_anticonformal}, which resembles the case of impulsive waves in a full \AdS spacetime \cite{PodGri97}.
Note that in the limit $a^2\rightarrow\infty$ (that is, when the cosmological constant and the electromagnetic field approach zero) all the metrics (\ref{generalwave}) become impulsive \pp waves.

The analysis of Secs.~\ref{sec_curvature}, \ref{sec_symmetries} and \ref{sec_radiative} can be straightforwardly adapted to any of the above possible backgrounds. 

\section{Concluding remarks}

A new class of exact solution of Einstein's equations with a positive cosmological constant has been presented. This describes impulsive gravitational and/or matter waves propagating in a Nariai universe, and thus completes the classification of non-expanding impulsive waves in spherically symmetric vacuum spacetimes. A convenient six-dimensional embedding formalism has been employed for the construction. The formal structure of the solutions has been shown to be similar to that of previously known non-expanding impulsive waves in (anti-)de~Sitter spacetimes \cite{HotTan93,PodGri98cqg}. Nevertheless, the background is now non-trivial and displays different topological properties. The geometrical approach has also been used for discussion of symmetries of these solution (following \cite{PodOrt01}) and for constructing impulsive waves in other direct product backgrounds, such as the anti-Nariai and the Bertotti--Robinson universe.

Vacuum field equations have been solved in full generality, and singularities of the solutions have been interpreted in terms of point multipole sources of gravitational waves. Our analysis followed the works by Griffiths and Podolsk\'y for solutions in Minkowski \cite{GriPod97} and (anti-)de~Sitter \cite{PodGri98cqg} backgrounds. But there is a difference. In the solutions \cite{GriPod97,PodGri98cqg}, the axially symmetric monopole terms are physically understood as fields generated by ultra-relativistic particles, initially obtained by an appropriate boosting technique by Aichelburg and Sexl \cite{AicSex71} and Hotta and Tanaka \cite{HotTan93}. A generalization  by Podolsk\'y and Griffiths themselves \cite{PodGri98prd} extends such an interpretation to higher multipole terms (at least when $\Lambda=0$). However, it seems that no exact static solutions are known in an asymptotically Nariai spacetime. Therefore, so far we can not relate the above null solutions to any such field boosted to the speed of light.

Finally, we have observed that (similarly as in the case of constant curvature spacetimes \cite{Podolsky98nonexp}) these impulsive solutions in the Nariai universe belong to a more general family of Kundt spacetimes. This family can thus be interpreted as representing exact sandwich gravitational waves of finite duration, or even the full Nariai cosmos filled with gravitational radiation. It seems to us that these solutions did not appear previously, at least in explicit form. Very similar metrics have been already considered, see, e.g., equations (27.54) and (31.37) in \cite{kramerbook} (for special choices of the arbitrary functions/parameters therein). However, these describe non-vacuum type-$N$ spacetimes. According to considerations in Secs.~\ref{sec_curvature} and \ref{sec_directproducts}, they can be interpreted as a Bertotti--Robinson universe with gravitational radiation.

\begin{acknowledgments}

I wish to thank Sergi Hildebrandt for bringing my attention to the Nariai spacetime. I am also grateful to Ji\v{r}\'{\i}~Podolsk\'y, Luciano Vanzo and Sergio Zerbini for a careful reading of the manuscript and useful suggestions.

\end{acknowledgments}

\appendix

\section{}
\label{app_multipoles}

By a slight modification of the derivation by Podolsk\'y and
Griffiths \cite{PodGri98cqg} for the case of impulsive waves in (anti-) de Sitter
spacetimes, we present here basic relations leading to
formulae (\ref{mpole}). One starts by expanding the function $\ln(1-z)^{1/2}$
for $z\in(-1,1)$ in terms of the complete system of Legendre polynomials as (see, e.g.,
\cite{hansenbook})
\begin{equation}
 \ln(1-z)^{1/2}={\textstyle
 \frac{1}{2}}(\ln2-1)-\sum_{l=1}^\infty\frac{l+\frac{1}{2}}{l(l+1)}P_l(z).
  \label{expansionln}
\end{equation}
Since $P_l(-z)=(-1)^lP_l(z)$, it also holds 
\begin{equation}
 \frac{1}{2}\ln\frac{1+z}{1-z}=\sum_{l=1}^\infty\frac{l+\frac{1}{2}}{l(l+1)}P_l(z)[1-(-1)^l] .
 \label{expansionQ00}
\end{equation}
For higher $m$-terms, combing the definition (\ref{Fm}) for
$F_{m}(z)$ with (\ref{expansionln}) and the recurrence formula for
the associated Legendre functions of the first kind,
\begin{equation}
P_l^m(z)=(-1)^m(1-z^2)^{m/2}\frac{\d^m}{\d z^m}P_l(z) ,
 \label{recurrencePlm}
\end{equation}
one gets
\begin{equation}
 F_m(z)=-(-1)^m\sum_{l=1}^\infty\frac{l+\frac{1}{2}}{l(l+1)}P_l^m(z)
 .
 \label{expansionFm}
\end{equation}
Recalling that $\int_{-1}^1\d z
P_l(z)P_j(z)=(l+1/2)^{-1}\delta_{lj}$ and $P_l(1)=1$, one can write the distributional expansion
\begin{equation}
 \delta(1-z)=\sum_{l=0}^\infty{\textstyle\left(l+\frac{1}{2}\right)}P_l(z) .
 \label{expansiondelta}
\end{equation}
If now the operator
${\bf L_m}\equiv\pa_z[(1-z^2)\pa_z]-m^2(1-z^2)^{-1}$ is introduced (for
$m\ge 0$), the identity ${\bf L_m}P_l^m(z)=-l(l+1)P_l^m(z)$ follows, since
$P_l^m(z)$ are solutions of an associated Legendre equation.
Applying such an identity to (\ref{expansionQ00}) and
(\ref{expansionFm}) and making use of (\ref{recurrencePlm}) and
(\ref{expansiondelta}), one gets \beqn
 & & {\bf L_0}\frac{1}{2}\ln\frac{1+z}{1-z}=\delta(1+z)-\delta(1-z) ,
 \nonumber \\
 & & {\bf L_m}F_m(z)=(1-z^2)^{m/2}\delta^{(m)}(1-z) ,
\eeqn where $\delta^{(m)}(1-z)\equiv\d^m\delta(1-z)/\d z^m$. An analogous proof can be carried out for
$F_{-m}(z)$.

\section{}
\label{app_energy}

The (distributional) energy conditions obeyed by the idealized point sources (\ref{mpole}) of Sec.~\ref{sec_curvature} are here discussed. 
First of all, for pure radiation matter ($T_{\mu\nu}\sim k_\mu k_\nu$) the weak, strong and dominant energy conditions turn out to be completely equivalent, due to the null character of $\mbox{\boldmath$k$}$. Then, it suffices to concentrate on the weak condition. Given a non-spacelike vector $\mbox{\boldmath$t$}$ and considering the results of Sec.~\ref{sec_curvature}, that reads 
\begin{equation}
T_{\mu\nu}t^\mu t^\nu=(32\pi)^{-1}\Lambda\,J(z,\phi)\,\delta(u)(\mbox{\boldmath$k$}\cdot\mbox{\boldmath$t$})^2\ge0 .
\label{weak}
\end{equation}
Clearly, this is automatically satisfied everywhere but at $u=0$, where the possible matter is localized (from now on, we also disregard the trivial case  $\delta(u)(\mbox{\boldmath$k$}\cdot\mbox{\boldmath$t$})^2=0$). On the wave front, the crucial sign is given by $J(z,\phi)$, since $\delta(u)$ is a non-negative distribution. Now, 
the multipoles with odd $m$  in (\ref{mpole}) have to be considered only as formal, since contain terms (such as $x^{m/2}\delta^{(m)}(x)$) which are not defined as distributions within Schwartz's theory. These will not be discussed here. The even-multipoles, instead, are described by distributions with a non-definite ``sign'', thus violating (\ref{weak}). 
This agrees with physical intuition as, roughly speaking, a multipole must contain ``charge'' densities of both signs. As expected, it turns out that the surface integral of such $J_{\pm m}$ is indeed vanishing (thus the total energy is non-negative).
On the other hand, the monopole $J_0$ in (\ref{mpole}) is described by the difference of two distributions with a well-defined sign. Then, it should be interpreted as representing two particles with equal and opposite energy densities. Note that the ``unphysical'' negative energy density is in this case mathematically unavoidable, and can geometrically be understood by considering the equivalent electrostatic problem of a point charge on a sphere (the electric lines of force generated by a source at the South pole must re-converge on a sink at the North pole, instead of spreading out at infinity, as in the planar case).

We conclude showing that, anyway, combined solutions (with point particles and null dust) exist which do not violate the energy conditions. We may simply adapt a ``heuristic argument'' by Balasin and Nachbagauer \cite{BalNac95} (who also indicate how to make the proof rigorous). Thanks to the generalized Kerr--Schild form of the metric (\ref{4-wave}) (see Footnote~\ref{note_KS}), the squared norm of $\mbox{\boldmath$t$}$ can be decomposed as 
\begin{equation}
 \mbox{\boldmath$t$}\cdot\mbox{\boldmath$t$}={\textstyle\frac{1}{2}}H\delta(u)(\mbox{\boldmath$k$}\cdot\mbox{\boldmath$t$})^2+\tilde{\mbox{\boldmath$g$}}(\mbox{\boldmath$t$},\mbox{\boldmath$t$})\le 0 . 
 \label{causality}
\end{equation}
In this, it is the first ``infinite'' term that determines $\mbox{\boldmath$t$}$ being causal. Therefore, if we consider a profile function $H$ which is strictly positive on the whole wave front ($z\in[-1,1]$), (\ref{causality}) can never hold there. In this case, (\ref{weak}) does not have to be satisfied, thus not restricting $J(z,\phi)$.
For instance, the positive function $H(z)=e^z-\ln(1-z^2)^{1/2}$ is associated with the source term $J(z)=e^z(z^2+2z-1)+\delta(1-z)+\delta(1+z)-1$.


\end{document}